\documentclass[pra,twoside,showpacs,twocolumn,floatfix]{revtex4}
\usepackage{times,graphicx,amsmath,bbm,mathrsfs,amssymb}
\newcommand{\ket}[1]{\vert #1 \rangle} 
\newcommand{\bra}[1]{\langle #1 \vert}
\newcommand{\bmsigma}{\boldsymbol \sigma}
\newcommand{\h}{{\rm h}}
\newcommand{\bfp}{\boldsymbol p}
\newcommand{\bfq}{\boldsymbol q}
\newcommand{\dmat}{{\varrho}}
\newcommand{\qobs}{{A}}

\newcommand{\MinKEnt}{mKE~}
\begin{document}
\title{Quantum estimation via minimum Kullback entropy principle}
\author{Stefano Olivares}\email{stefano.olivares@mi.infn.it}
\affiliation{Dipartimento di Fisica dell'Universit\`a di Milano, I-20133, Italia}
\affiliation{Dipartimento di Matematica e Fisica dell'Universit\`a
dell'Insubria, Como, Italia}
\author{Matteo G. A. Paris}\email{matteo.paris@unimi.it}
\affiliation{Dipartimento di Fisica dell'Universit\`a di Milano, I-20133, Italia}
\affiliation{Institute for Scientific Interchange Foundation, I-10133 Torino,
Italia}
\date{\today}
\begin{abstract}
We address quantum estimation in situations where one has at
disposal data from the measurement of an incomplete set of
observables and some {\em a priori} information on the state
itself. By expressing the {\em a priori} information in terms of
a bias toward a given state, the problem may be faced by
minimizing the quantum relative entropy (Kullback entropy) with
the constraint of reproducing the data.  We exploit the resulting
{\em minimum Kullback entropy principle} for the estimation of a
quantum state from the measurement of a single observable, either
from the sole mean value or from the complete probability
distribution, and apply it as a tool for the estimation of {\em
weak} Hamiltonian processes. Qubit and harmonic oscillator
systems are analyzed in some details.
\end{abstract}
\maketitle
\section{Introduction}\label{s:intro}
Quantum estimation of states and operations is a relevant topic
in the broad field of quantum information science \cite{LNP649}.
The subject has a fundamental interest of its own, since it
concerns the characterization of the the basic objects of the
quantum description of physical systems.  In addition, quantum
estimation techniques have been receiving attention for their
role in the characterization of gates and registers at the
quantum level, which itself is a basic tool in the development of
quantum information technology. 
\par
In order to characterize a quantum system one may measure an
observable, or a set of observables, on repeated preparations of
the system.  As a matter of fact, the set of observable is
usually {\em incomplete}, i.e., it is not sufficient to give
a complete quantum information on the system. In other words, it
is not possible to deterministically reconstruct the full density
matrix of the system from the measured data \cite{gentomo}. In
these cases, the question is not that of finding the actual state
of the system, but rather that of estimating the state that best
represents the knowledge we have acquired about the system from
the measured data \cite{BuzLNP}. If we assume not to have any
prior information about the system, and quantify this ignorance
by entropy, than the best estimate may be found by the Jaynes
maximum-entropy principle (MaxEnt) \cite{maxent:57}, which
include the information obtained by measurements while not
allowing one to draw any conclusions not warranted by the data
themselves.
\par
On the other hand, in most cases one has some {\em a priori}
information on the state of system under investigation. This may
come, e.g., by some energy constraint or by the consideration
that the system has been weakly perturbed from an initial given
preparation which is under control of the experimenter. A
question naturally arises on how this {\em a priori} information
may be incorporated into the estimation procedure and whether
this can be done together with constraint of reproducing the
observed data. The quantum mechanical way to incorporate some
{\em a priori} information is that of bias of the state to be
estimated towards a given quantum state $\tau$, which expresses
the information at hand of the experimenter and which may contain
one or more free parameters depending on the amount of {\em a
priori} information at disposal. Upon quantifying the bias
towards $\tau$ through relative (Kullback) entropy
\cite{umega,vedral02} an estimate may be found by minimizing this
quantity with the constraint of reproducing the data.  This
minimum Kullback-entropy principle (mKE) emerges naturally as a
way to estimate the quantum state of a system from incomplete
data, when some {\em a priori} information on the state is
available, i.e., when a bias towards a known quantum state is
present. We will exploit this ideas to estimate the quantum state
of a system from the measurement of a single observable, either
from the sole mean value or from the full probability
distribution, and apply it to the estimation of {\em weak}
Hamiltonian processes. In particular, the case of a weak, but
otherwise generic, Hamiltonian is analyzed in some details, with
emphasis to qubit and harmonic oscillator systems.
\par
The paper is structured as follows. In Section \ref{s:intro2} we
review classical Kullback-Leibler divergence for probability
distribution and its quantum counterpart, the quantum relative
Kullback entropy.  In addition, we state the minimum Kullback
entropy principle which is used in the subsequent Sections.  In
Section \ref{s:st:MKE} we exploit \MinKEnt for state estimation.
In particular, we analyze \MinKEnt estimation from the
measurement of a single observable, either from the sole mean
value or from the complete probability distribution.  Qubit and
harmonic oscillator systems are analyzed in some details.  In
Section \ref{s:wh:MKE} we address \MinKEnt as a tool in order to
estimate a weak Hamiltonian through suitable measurements
performed on the evolved states. Section \ref{s:conc} closes the
paper with some concluding remarks.
\section{Kullback-Leibler divergence and quantum Kullback entropy}
\label{s:intro2}
Let us consider a {\em classical} system which can be in any of
$N$ states and let $p_k$ be the probability for the $k$-th state.
Suppose we want to estimate the probability distribution $\bfp =
\{ p_k \}$. If all that we know is the number of possible states,
there is no way to do better than choosing $p_k = 1/N$, $\forall
k$. Fortunately, in general we have further information about our
system, e.g., we have at disposal the value of certain functions
of the state of the system. Thus, we can use these as {\em
constraints} to estimate $\bfp$. The problem of finding the most
likely distribution satisfying a set of given constraints was
solved by Jaynes, who proposed a general method of inference
known as principle of maximum entropy (MaxEnt) \cite{maxent:57}.
According to MaxEnt principle the best estimate of $\bfp$ is that
{\em maximizing} the Shannon entropy
\begin{equation}\label{shannon}
H(\bfp ) = - \sum_k p_k \log p_k\,
\end{equation}
given the constraints.  As it was described by Jaynes, such a
distribution is the one that {\em agrees with what is known, but
expresses maximum uncertainty} with respect to all other matters
\cite{cybe:1958}.
\par
On the other hand, it is not unlikely that the probability
distribution we want to estimate is biased towards a {\em prior}
distribution $\bfq = \{ q_k\}$. In this case it is useful to
introduce the Kullback entropy \cite{kull51,kullent:59}:
\begin{equation}\label{kullback}
K(\bfp |\bfq ) = \sum_k p_k \log(p_k/q_k),
\end{equation}
that represents the relative entropy of a {\em posterior}
distribution $\bfp$ relative to the {\em prior} $\bfq$. It is
worth noting that in the absence of any knowledge on $\bfq$,
i.e., $q_k=1/N$ (uniform distribution), we have
$K(\bfp|\bfq)=\log N - H(\bfp )$: this is just the opposite of
the Shannon entropy up to an additive constant. On the contrary,
when $\bfq$ is accessible, the Kullback entropy (\ref{kullback})
leads to a generalized method of inference: the most likely
distribution $\bfp$ with a bias towards $\bfq$ is the one {\em
minimizing} $K(\bfp |\bfq)$ given the constraints. This is the
classical principle of minimum Kullback entropy (mKE), also
referred to as the principle of minimum relative entropy
\cite{kull51,kullent:59,k1}, which found applications in several
branches of science \cite{k2,k3,k4}.  Indeed, minimizing the
relative entropy has all the important attributes of the maximum
entropy approach with the advantage that prior information may be
easily included.
\par
In the following we will deal with the quantum version of the
\MinKEnt principle and its application to quantum estimation of
states and operations. Before going to the main point let us
spend few words on how this works in the classical case.  Let us
consider the problem of finding the posterior $\bfp$ given the
prior $\bfq$ and the constraints $\sum_k p_k=1$ (normalization)
and $\sum_k p_k A_k=\langle A \rangle$, i.e., we assume to
know the first moment of the quantity $A$. Using the Lagrange
multipliers method the problem reduces to the minimization of
\begin{align}
F(\bfp,\lambda,\eta)= K(\bfp |\bfq ) &+\eta\,\left( \sum_k p_k 
- 1 \right) \nonumber \\ &+ 
\lambda\,\left(\sum_k p_k A_k-\langle A \rangle\right),
\end{align}
$\eta$ and $\lambda$ being Lagrange multipliers. We have:
\begin{align}
\label{systa}
0& =\log(p_k/q_k) + 1 + A_k \lambda + \eta \quad (k=1,\dots,N) \\ 
\label{systb} 1&=\sum_k p_k \\
\label{systc} \langle A \rangle & = \sum_k p_k A_k\,.
\end{align}
Solving the first equation with respect to $p_k$ and using the condition
(\ref{systb}) we obtain:
\begin{equation} \label{pkt:class}
p_k(\lambda) = \frac{q_k\, e^{-A_k \lambda}}{\sum_s 
q_s \, e^{-A_s \lambda}}\,,
\end{equation}
where we explicitly wrote the dependence of $p_k$ on the Lagrange
multiplier $\lambda$. Note that for $\lambda=0$ one has $p_k =
q_k$.  Differentiating Eq.~(\ref{pkt:class}) and using the
constraint (\ref{systc}) we obtain the differential equation:
\begin{equation}\label{class:traj}
\frac{d p_k}{d \lambda} = -(A_k - \langle A \rangle)\, p_k.
\end{equation}
Upon considering the distribution $\bfp$ as a point in the
probability distribution simplex, then Eq.~(\ref{class:traj}) can
be seen as a trajectory in such a space. According to the
Lagrange multipliers method, if we integrate the trajectory
(\ref{class:traj}) assuming $\bfp(0)=\bfq$ for $\lambda=0$, then the
minimum of Kullback entropy is achieved when the trajectory
passes through the surface satisfying the constraint $\sum_k p_k
A_k=\langle A \rangle$ \cite{Braun:PLA:96}.  
\subsection{Minimum Kullback entropy principle}
Let us turn our attention to quantum mechanics.  The problem is
now to find the most appropriate density matrix $\dmat$
describing a quantum system that should satisfy some given
constraints, which, in turn express  the results of an incomplete
set of measurements. If there is no a priori information, then
the optimal choice is given by the density matrix $\dmat$
which maximizes the von Neumann entropy
\begin{equation}
H (\dmat) = - {\rm Tr}[\dmat \log \dmat]\:,
\end{equation}
and satisfies the constraints, i.e., reproduces the observed
data. This is the quantum counterpart of the MaxEnt principle
introduced above \cite{BuzLNP}.  On the other hand, when we have
some a priori information about the system under investigation
then the state to be estimated has a bias towards a prior one
$\tau$. This is the case, for example, of a quantum system
evolved according to an unknown but weak Hamiltonian starting
from a known initial state. In order to build a proper estimation
scheme for these situations let us consider the quantum Kullback
entropy, defined as follows \cite{umega,vedral02}:
\begin{equation}\label{q:kull}
K (\dmat|\tau) = {\rm Tr}\left[
\dmat (\log \dmat - \log \tau) \right],
\end{equation}
that is the relative quantum entropy of the density matrix
$\dmat$ with respect to $\tau$.
\par
As well as in the classical case a probability distribution can
be seen as a point, in the quantum case the role of ``point'' is
played by the density matrix $\dmat$. We can also associate a
family of surfaces ${\rm Tr}[\qobs \dmat] = \langle \qobs
\rangle$ to  each quantum observable $\qobs$. Moreover, it is
still possible to define a suitable metric in the space of the
density matrices (the Hilbert space) \cite{Braun:PRL:94}: this
allows us to introduce the infinitesimal increment $d \dmat$.  In
this way one can obtain the following equation for the quantum
trajectory \cite{Braun:PLA:96}:
\begin{equation}\label{q:traj}
\frac{d \dmat}{d \lambda} =
-\frac12 \left\{ \dmat, \qobs - \langle \qobs \rangle \right\},
\end{equation}
where $\{ \hat X, \hat Y\} = \hat X \hat Y + \hat Y \hat X$ and
$\lambda$ is again a Lagrange multiplier.  Eq.~(\ref{q:traj}) is
the quantum counterpart of Eq.~(\ref{class:traj}) and satisfies
the constraints ${\rm Tr}[\dmat]=1$ and ${\rm Tr}[\qobs \dmat] =
\langle \qobs \rangle$, i.e., we are assuming that the
expectation value $\langle \qobs \rangle$ of $\qobs$ is known. By
formal integration of Eq.~(\ref{q:traj}), it is straightforward
to obtain the solution
\begin{equation}\label{kull:infer}
\dmat(\lambda) =
\frac{e^{-\qobs \lambda /2} \tau\, e^{-\qobs \lambda /2}}
{{\rm Tr}\left[\tau\,  e^{-\qobs \lambda}\right]},
\end{equation}
where we assumed that the formal integration starts from
$\dmat(0)=\tau$, $\tau$ being the prior density matrix.
\par
There are cases in which the trajectory (\ref{q:traj}) yields to
the optimal state with respect to the quantum Kullback entropy,
i.e., $K (\dmat(\lambda)|\tau)$ is minimized \cite{Braun:PRL:94}.
In this paper we focus our attention precisely on one of these
cases, namely, when the prior $\tau$ and the posterior $\dmat$
are {\em close} each other according to the Fisher 
information metric.
\subsection{Remarks}
As it may be expected, when the prior information is very weak 
mKE principle reduces to MaxEnt one. On the other hand, some 
prior knowledge is often present and mKE principle fully
exploit the additional information to improve the estimation
procedure. A question may arise on the choice of the 
measurements: as a general rule those should be tuned in order 
to add information with respect to the prior being otherwise 
useless for estimation purposes. In other words, the information 
coming from the measurements should not be subsumed by the prior
one.
\par 
Kullback-Leibler divergence and quantum Kullback entropy are
involved in different aspects of quantum estimation of states and
operations, including assessments of priors \cite{sla} in the
context of quantum Bayes rule \cite{bay}.  Here we briefly
mention few relevant applications in order to enlighten analogies 
and differences with our approach.  
\par 
Let us first consider a situation in which we have no a priori
information and want to estimate the state of a system from the
measurement of a set of a set of projector
$A_k=|\varphi_k\rangle\langle\varphi_k |$.  In this case the
maximum-likelihood estimate of the state \cite{mlik} is the
density matrix $\varrho$ that minimizes the Kullback-Leibler
divergence $\sum_k p_k \log (p_k/\varrho_k)$ between the observed
probability distribution and the quantum mechanical prediction
$\varrho_k=\langle \varphi_k | \varrho | \varphi_k\rangle$
\cite{suha}.  In other words, maximum-likelihood estimation leads
to the state that fits the given data obtained by the given
measurement without using prior information about the quantum
state.  
\par 
In Ref.~\cite{Kloss} Kullback entropy is used for quantum
estimation as a {\em loss} (cost) function in the search of 
the optimal predictive density matrix in a Bayesian (global)
approach. In other words, best estimate is found by minimizing
the average Kullback entropy with respect to the true state.
In the same perspective, Kullback entropy has been also used
as a regularizing functional in seeking solutions to
multivariable and multidimensional moment problems \cite{geo}. 
\par
Finally, notice that symmetrized version of the Kullback entropy
has been also suggested \cite{jsd} and may be employed to assess
the distance between quantum states. 
\section{Minimum Kullback principle for state estimation}
\label{s:st:MKE}
In this section we exploit \MinKEnt principle for the estimation
of the full density matrix $\dmat$ from an incomplete set of
measurements.  We assume that the system under investigation has
a bias towards the known prior $\tau$ and first consider to have
access to the mean value of a single observable. Then the
analysis is generalized to the case of $N$ observables, with
application to the measurement of the complete distribution of a
single observable.
\subsection{Measurement of a single observable}
In this Section we assume that only the mean value of the
observable $\qobs$ can be measured. This is the simplest
observation level \cite{buz:AnnPh:98} one may devise for a
quantum system, and is generally considered to provide only a
little amount of information about the state under investigation.
Upon applying the quantum \MinKEnt principle, the best estimate
for the density matrix, compatible with the bias is given by
Eq.~(\ref{kull:infer}). By introducing the partition function
$Z={\rm Tr}[\tau e^{-\qobs \lambda}]$ the estimated density
matrix reads as follows
\begin{equation}\label{MKE:Braun}
\dmat = \frac{1}{Z}\,e^{-\frac12 \qobs \lambda}\, \tau\, 
e^{-\frac12\qobs \lambda}.
\end{equation}
Moreover, using the spe\-ctral de\-compo\-sition $\qobs = \sum_k
\alpha_k \ket{\varphi_k}\bra{\varphi_k}$, $\{\ket{\varphi_k}\}$
being a complete orthonormal system of eigenvectors of the
operator $\qobs$, we can write: $\exp\{-\frac12 \qobs \lambda\} =
\sum_k \exp\{-\frac12 \alpha_k \lambda\}$ so that
\begin{align}\label{Zd}
Z & = \sum_k e^{-\alpha_k\lambda} \bra{\varphi_k}\tau\ket{\varphi_k} \\
\bra{\varphi_m} \varrho \ket{\varphi_n} & = \frac1Z 
e^{-\frac12 (\alpha_m + \alpha_n )\lambda} 
\bra{\varphi_m} \tau \ket{\varphi_n}\:. \label{p:k}
\end{align}
In this way, given the initial density matrix $\tau$ 
it is possible to estimate the complete state $\dmat$
as follows
\begin{align}
\dmat & = \frac1Z \sum_{n,m} \bra{\varphi_m}\tau\ket{\varphi_n}\, 
e^{-\frac12 (\alpha_n +\alpha_m)\lambda}\, \ket{\varphi_m}
\bra{\varphi_n}\,, \label{best:est}
\end{align}
where the value of the Lagrange multiplier $\lambda$ is obtained
as a (numerical) solution of the equation
$\hbox{Tr}\left[\varrho\, A\right] = \langle A \rangle$, i.e.,
\begin{align}
\langle A \rangle = \frac{ \sum_k \bra{\varphi_k}\tau\ket{\varphi_k} 
e^{-\alpha_k\lambda} \alpha_k}{\sum_k \bra{\varphi_k}\tau\ket{\varphi_k} 
e^{-\alpha_k\lambda}}\:,
\end{align}
which, of course, is equivalent to the relation
$-\partial_\lambda \log Z = \langle A \rangle$.  It is worth
noting that the estimate (\ref{best:est}) for the density matrix
has support in the same subspace of Hilbert space where $\tau$
does.  Moreover, if $\tau = \ket{\varphi_{\overline{n}}}
\bra{\varphi_{\overline{n}}}$, i.e., if the prior density matrix is a
projector onto the subspace generated by the eigenvector
$\ket{\varphi_{\overline{n}}}$ of $\qobs$, then $ \bra{\varphi_n}
\tau \ket{\varphi_n} =\delta_{n\overline{n}}$ and the \MinKEnt
principle reduces to the MaxEnt one, as in the classical case.
\subsection{Measurement of $\boldsymbol N$ observables}
\label{s:N:obs}
Here we assume that the mean values of $N$ different observables
$\qobs_k$, $k=1,\ldots, N$ are experimentally accessible.  This
means that we we have $N$ constraints $\langle \qobs_k \rangle$,
to be considered. Upon writing the trajectory (\ref{q:traj}) for
each constraint, we have a system of $N$ differential equations
which can be written in the following compact form:
\begin{equation}
\sum_{k=1}^{N} \frac{d \dmat}{d \lambda_k} = -\frac12
\left\{ \varrho,
\sum_{k=1}^{N} \left( \qobs_k - \langle \qobs_k \rangle\right)\right\} ,
\end{equation}
The solution can be written as:
\begin{equation}\label{nmke}
\dmat = \frac{1}{Z} e^{-\frac12 \sum_k \qobs_k \lambda_k}
\,\tau\, e^{-\frac12 \sum_k \qobs_k \lambda_k}
\end{equation}
where $\lambda_1,\ldots, \lambda_N$ are Lagrange multipliers, and
we assumed $\dmat(0) = \tau$.  The partition function $Z$ reads:
\begin{equation}
\label{Zn}
Z = {\rm Tr} \left[\tau\, e^{-\sum_k \qobs_k \lambda_k}
\right]\,.
\end{equation}
Again, the values of the multipliers $\lambda_k$ are obtained
solving the system of equations
\begin{align}
\label{Cn}
{\rm Tr}[\dmat\,\qobs_k] = \langle \qobs_k \rangle 
\qquad k=1,\ldots,N\:.
\end{align}
The above analysis allows us to apply \MinKEnt principle also
starting from the measurement of the full distributions of an
observable.  In this case the set of observables to be taken into
account are the orthogonal (commuting) eigenprojectors $A_k =
\ket{\varphi_k} \bra{\varphi_k}$, $\langle \varphi_k|\varphi_s
\rangle=\delta_{ks}$ of the measured observables. The constraints
$\hbox{Tr}\left[\varrho\,A_k\right] = p_k$, correspond to the
measured distribution.  Eq. (\ref{Zn}) and (\ref{Cn}) rewrite as 
\begin{align}\label{xxy}
Z & = \sum_k e^{-\lambda_k}\, \langle \varphi_k|\tau |
\varphi_k \rangle \\
p_k & = \frac1Z e^{-\lambda_k} \langle \varphi_k|\tau |
\varphi_k \rangle\:. 
\label{xxx}
\end{align}
Finally, taking matrix elements of (\ref{nmke}) and using
(\ref{xxy}) and (\ref{xxx}) it is possible to reconstruct the
posterior state, given the initial density matrix $\tau$ and the
measured probabilities $p_k$
\begin{equation}\label{best:est:pn}
\dmat = \sum_{n,m}
\frac{
\bra{\varphi_m}\tau\ket{\varphi_n}}{\sqrt{
\bra{\varphi_m}\tau\ket{\varphi_m}
\bra{\varphi_n}\tau\ket{\varphi_n}}}\, 
\sqrt{p_m\,p_n}\:\ket{\varphi_m}\bra{\varphi_n}.
\end{equation}
Notice that in this case we have been able to back-substituting the 
Lagrange multipliers, i.e., Eq.~(\ref{best:est:pn}) no longer 
depends on the $\lambda_k$'s.
\subsection{The qubit case}
Here we address the estimation of a qubit state starting from the
measurement of a single observable \cite{2d}. In order to apply
the \MinKEnt principle we assume to have a bias towards the state
$\tau$ and choose an observable to measure on the system. The
measured quantity is the spin along direction $\vec{n}$, which is
described by the operator: 
\begin{equation}\label{qubit:meas}
\qobs = \vec{n}\cdot \vec{\bmsigma},
\end{equation}
where we defined the vector
$\vec{\bmsigma} = (\bmsigma_1,\bmsigma_2,\bmsigma_3)$,
$\bmsigma_k$, $k=1,2,3$, being the Pauli matrices. 
Upon writing the  prior state in the Pauli basis 
\begin{equation}
\tau = \frac12 (\mathbbm{1} + \vec{\tau}\cdot \vec{\bmsigma}),
\quad |\vec \tau| \le 1\,,
\end{equation}
Eq.~(\ref{MKE:Braun}) reads:
\begin{equation}
\dmat = \frac12 (\mathbbm{1} + \vec{v}\cdot \vec{\bmsigma}),
\end{equation}
with
\begin{equation}\label{qubit:est}
\vec{v} = \frac{\vec{\tau} +
2 \sinh^2(\lambda/2) \left(\vec{n}\cdot\vec{\tau}\right) \vec{n}
- \sinh \lambda\, \vec{n}}
{\cosh \lambda - \vec{\tau}\cdot\vec{n}\,\sinh \lambda},
\end{equation}
where we used $Z=\cosh \lambda - \vec{\tau}\cdot\vec{n}\,\sinh \lambda$.
Now, thanks to the constraint:
\begin{equation}\label{con:qubit}
{\rm Tr}[\dmat\,\vec{n}\cdot \vec{\bmsigma}]=
\langle \vec{n}\cdot \vec{\bmsigma} \rangle,
\end{equation}
we can calculate the value of the Lagrange multiplier $\lambda$, obtaining:
\begin{equation}\label{lamqub}
\lambda = {\rm arctanh}\: 
\frac{\vec{\tau}\cdot\vec{n} - \langle \vec{n}\cdot \vec{\bmsigma} \rangle}
{1- \langle \vec{n}\cdot \vec{\bmsigma} \rangle (\vec{\tau}\cdot\vec{n})}\,.
\end{equation}
In order to have a compact expression of the result let us consider an 
operator basis composed by spin operators along three orthogonal directions
$\vec{n}_1\perp\vec{n}_2\perp\vec{n}_3$, with $\vec{n}_1\equiv\vec{n}$.
In this way we can express the components of the vector $\vec{v}$ as follows:
\begin{align}\label{qubit:3comp}
\vec{v}\cdot\vec{n}_1 & = \langle \vec{n}\cdot \vec{\bmsigma} \rangle  \\ 
\vec{v}\cdot\vec{n}_k & = \vec{\tau}\cdot\vec{n}_k\,
\sqrt{\frac{1-\langle \vec{n}\cdot \vec{\bmsigma} \rangle^2}
{1-(\vec{\tau}\cdot\vec{n})^2}} \quad (k=2,3)\label{qubit:3comp:bis}\,.
\end{align}
Eqs. (\ref{qubit:3comp}) says that the estimated 
Bloch component in the direction of the measured observable is equal to the
measured mean value, whereas the two other orthogonal components are 
obtained from the prior one by a common shrinking factor.
\par
As an example, let us assume $\vec{n} = (1,0,0)$ and $\vec{\tau} = (0,0,1)$:
this is the case of the measurement of $\bmsigma_1$ (spin along $x$
direction) and bias towards $+z$ direction. We have
$x = -\langle \bmsigma_1\rangle$ and, then,
\begin{equation}
\vec v = \left(\langle \bmsigma_1\rangle , 0 ,
\sqrt{1-\langle \bmsigma_1\rangle^2} \right),
\end{equation}
which satisfies Eqs.~(\ref{qubit:3comp}) and (\ref{qubit:3comp:bis}).
\subsection{The harmonic oscillator case}
Now we face the problem of estimating the state of a harmonic
oscillator with a bias towards a coherent state 
$\tau = \ket{\alpha}\bra{\alpha}$, with $\alpha\in\mathbb{C}$. 
Let us consider a photon number measurement, i.e., if $a$ denotes 
the annihilation operator, $[a,a^\dag]=1$, the observable expressed by 
$\qobs \equiv a^{\dag} a = \sum_n n \ket{n}\bra{n}$, $\{ \ket{n} \}$
being the photon number basis. Since
\begin{equation}
\bra{n} \tau \ket{m} = 
\frac{\alpha^n\bar\alpha^m \, e^{- |\alpha^2|}}{\sqrt{n! m!}},
\end{equation}
we have that the \MinKEnt estimated state is still a
coherent state with amplitude
\begin{equation}
\beta = \alpha\, e^{-\lambda/2}\,.
\end{equation}
Now, using the constraint
\begin{equation}
{\rm Tr}[\dmat\,a^\dag a] = N,
\end{equation}
with $N$ obtained from the experiment, we can evaluate the value
of the Lagrange multiplier $\lambda$, namely,
\begin{equation}
\lambda = \log(|\alpha|^2 / N)\:,
\end{equation}
also obtaining $Z=\exp \{N\}$. Finally, upon substituting in (\ref{best:est})  
we arrive at
\begin{align}
\varrho &= e^{-N} \sum_{nm} \left(N/|\alpha|^2\right)^{(n+m)/2} 
\frac{\alpha^n\bar\alpha^m}{\sqrt{n! m!}} \nonumber \\ 
&\equiv |\sqrt{N}e^{i \phi} \rangle\langle \sqrt{N} e^{i\phi}|\,,
\end{align}
with $\phi=\arg \alpha$. In other words, the best estimate
according to \MinKEnt is a coherent state with average number of photons
equal to the measured one and phase equal to that of the prior coherent state.
Notice that the best estimate obtained using the MaxEnt principle
with the same constraint on the average number of photons, but without
the bias, would have been a thermal state with $N$ thermal photons.
\par
If  by some means the complete photon distribution $p_n$ is available,
the reconstructed state, given by Eq.~(\ref{best:est:pn}),
reads as follows (we assumed that the bias is still towards
$\tau=\ket{\alpha}\bra{\alpha}$, $\alpha\in\mathbbm C$):
\begin{equation}\label{coh:rec}
\dmat = \sum_{n,m} \sqrt{p_{n}\, p_{m}}\,e^{i\phi (n-m)} \ket{n}\bra{m}\:.
\end{equation}
Remarkably, Eq. (\ref{coh:rec}) no longer depends on the amplitude of
the prior state, which enters only trough the phase $\phi$. This makes
the above scheme quite promising though measuring the photon distribution 
is, in general, a challenging task. On the other hand, in the optical case it
is possible to reconstruct the $p_n$ by means of on/off photodetection
and maximum likelihood algorithm \cite{mog,ARR:PRA:70}, a method that has 
been recently verified in laboratory \cite{MB:PRL:95,MG:LP:06}.
Multichannel fiber loop detector \cite{olom,kb} may be also used.
\par
Finally, we notice that a special case of bias is that toward a 
Gaussian state. In fact the Kullback relative entropy of a state 
$\varrho$ with respect to a Gaussian state $\tau$, {\em with the same
covariance matrix} reduces to the difference of the Von Neumann
entropies \cite{rev} 
$$K (\dmat|\tau) = H(\tau) - H(\varrho)$$
and thus the \MinKEnt principle reduces to the MaxEnt. Notice, however, 
that this is not in contrast with the results above, since in that case 
the results of the measurement do not impose the equality of the 
covariance matrices. 
\section{mKE estimation of weak Hamiltonians}
\label{s:wh:MKE}
In the previous section we have shown how to fully reconstruct the
density matrix of a quantum system from incomplete data. Here we will 
see how it is possible to estimate a weak Hamiltonian $H$ by means of 
the \MinKEnt and suitable measurements onto the evolved states.
The idea behind this method is that, upon considering weak Hamiltonian
processes, the evolved state is not too different from the initial
one, i.e., it has  a natural bias towards the unperturbed state.
This allows one to estimate the parameters (matrix elements) of the
Hamiltonian from data obtained by an incomplete set of measurements on 
the evolved state, i.e., to use \MinKEnt principle as a effective 
tool for process estimation.
\subsection{The qubit case}
The initial state $\tau$ and the evolved state of the qubit
systems under investigation are connected by the transformation
\begin{equation}\label{time:ev}
\dmat_t = e^{-i H t} \tau\, e^{i H t}
\end{equation}
In. Eq.~(\ref{time:ev}) $t$ is the time evolution.
Using the Pauli basis, we can express the initial state
and the Hamiltonian as  
\begin{equation}
\tau = \frac12 \left( \mathbbm{1}
+ \vec{\tau}\cdot\vec{\bmsigma} \right),\quad |\vec{\tau}|\le 1, 
\end{equation}
and
\begin{equation}
H = \sum_{\nu=0}^{3} \h_\nu\, \bmsigma_\nu, \quad |\vec{\h}| \ll 1,
\end{equation}
respectively, where $\bmsigma_\nu$ are the Pauli matrices with
$\bmsigma_0 = \mathbbm{1}$, and $\vec{\bmsigma}$ has been defined above.
Expanding at the first order in $\vec{\h}$ Eq.~(\ref{time:ev}),
we obtain:
\begin{equation}\label{expansion}
\dmat_t = \tau + it [\tau,H] + o(|\vec{\h}|^2),
\end{equation}
with
\begin{align}
[\tau,H] = i \sum_{k,s=1}^{3} \tau_s\, \h_k\, \varepsilon_{ksl}\, 
\bmsigma_l,
\end{align}
$\varepsilon_{ksl}$ being the totally antisymmetric tensor,
$\varepsilon_{123}=1$. In this way, expansion (\ref{expansion}) can
be written as follows:
\begin{equation}
\dmat_t = \frac12 \left( \mathbbm{1}
+ \vec{w}\cdot\vec{\bmsigma} \right),
\end{equation}
where the Bloch vector is given by 
\begin{equation}\label{w:qubit}
\vec{w} \equiv \vec{\tau} + 2\, \vec{\h}\times \vec{\tau}\,.
\end{equation}
Eq.~(\ref{w:qubit}) represent a system of equation for the
unknowns $\vec{\h}$. The transfer matrix is singular, but the
system may be anyway inverted using the Moore-Penrose generalized
inverse \cite{MP:pinv}, thus leading to the expression
\begin{equation}\label{w:qubit1}
\vec{\h} =
\frac{\vec{\tau}\times \vec{w}}{2 |\vec{\tau}|^2} ,
\end{equation}
which provides an estimate for the Hamiltonian $\vec{\h}$ once an
estimate for $\vec{w}$ is given. The latter is obtained upon the
measurement of a spin observable described by the operator $\vec
n \cdot \vec \bmsigma$ onto evolved state $\dmat_t$.  The best
estimate for $\vec w$ according to \MinKEnt principle is given by
Eq.~(\ref{qubit:est}).  By substituting in Eq.~(\ref{w:qubit1})
and using (\ref{lamqub}) an estimate for $\vec{\h}$ is achieved:
\begin{align}\label{hq}
\vec{\h} = \frac{1}{2|\vec{\tau}|^2}\frac{(1-\sqrt{1-\kappa^2})
\vec{\tau}\cdot\vec{n}-\kappa}{1- \vec{\tau}\cdot\vec{n}\:\kappa}\:
\vec{\tau}\times\vec{n}\, 
\end{align}
where 
\begin{equation}\label{kappa}
\kappa = 
\frac{\vec{\tau}\cdot\vec{n} - \langle \vec{n}\cdot \vec{\bmsigma} \rangle}
{1- \langle \vec{n}\cdot \vec{\bmsigma} \rangle (\vec{\tau}\cdot\vec{n})}\,.
\end{equation}
As it is apparent
from Eq.~(\ref{hq}) the \MinKEnt estimate for the Hamiltonian
Bloch vector is orthogonal to both the prior state one and the
measurement direction. By a repeated randomized choice of the
measurement an effective reconstruction may be achieved for any,
but weak, qubit Hamiltonian. Generalizations to higher dimensional
systems may be also designed.
\subsection{The harmonic oscillator case}
Let us now assume that the expansion
\begin{equation}\label{expansion:HS}
\dmat (t) = \tau + it [\tau,H] + o(|H|^2),
\end{equation}
refers to the state of a harmonic oscillator evolving under the action
of a weak Hamiltonian $H$ that we want to estimate. We also assume that 
the full distribution of a single observable can be measured of the evolved
state. Therefore, the evolved density matrix $\varrho_t$ may be reconstructed 
by mKE starting from the observation level $\qobs_k =
\ket{\varphi_k}\bra{\varphi_k}$,
as described in Section \ref{s:N:obs}, and thus obtaining the 
state (\ref{best:est:pn}).
Using the basis $\{ \ket{\varphi_k} \}$ 
we can write Eq.~(\ref{expansion:HS}) as follows:
\begin{equation}\label{expansion:HS:nm}
\dmat_{mn}(t) =  \tau_{mn} + i t \sum_s 
\left ( H_{ms}\tau_{sn}  + 
\tau_{ms} H_{sn} \right) 
\end{equation}
where we used $H_{mn} \equiv \bra{\varphi_m} H \ket{\varphi_n}$.
Using the \MinKEnt estimate (\ref{best:est:pn}) for the evolved density
matrix we obtain the following hierarchy of equations
\begin{align}
\sum_s \big( H_{ms}\tau_{sn}   & + 
\tau_{ms} H_{sn} \big)  = \nonumber \\  & \frac{i}{t}
\tau_{mn} \left( 1- \sqrt{\frac{p_n p_m}{ 
\bra{\varphi_n}\tau\ket{\varphi_n}
\bra{\varphi_m}\tau\ket{\varphi_m}
}}\right)
\end{align}
where $p_n$ are the measured probabilities (the constraints used for
the mKE) and the matrix elements $H_{nm}$ are the unknowns.
\par
A relevant example, in which the Hamiltonian can be effectively
estimated using mKE, is that corresponding to $H = (g a + {\rm
h.c.})$, $a$ being the annihilation operator starting from the
sole measurement of the photon distributions.  The evolution
imposed by the Hamiltonian $H$ corresponds to the unitary
displacement operator $D(\beta) = \exp( \beta a^\dag - \beta^*
a)$, $\beta= g t$.  The problem is then to estimate the
displacement amplitude $\beta$ from the measured photon
distribution. We assume that the initial state is a coherent
state $\dmat (\alpha) = \ket{\alpha}\bra{\alpha}$.  For the sake
of simplicity we take $\alpha$ and $\beta$ as real.  Using the
photon number basis the evolved state may be written as 
\begin{equation}
\dmat (\beta) = e^{(\alpha + \beta)^2}\sum_{n,m}
\frac{(\alpha + \beta)^{n+m}}{\sqrt{n!\,m!}}\, \ket{n}\bra{m}\:.
\end{equation}
Assuming that the measurement of the photon number is made on the
evolved state and that \MinKEnt principle is used to estimate the
density matrix we equate the above expression to that given  in
Eq.~(\ref{coh:rec}) for $\phi=0$ (recall that $\alpha$ and
$\beta$ are taken as real). We thus obtain the following set of
equations:
\begin{equation}\label{coh:disp}
-(\alpha + \beta)^2 + (n+m)\, \ln (\alpha + \beta) =
\ln \sqrt{n!\,m!\,p_n\,p_m},
\end{equation}
to be solved for $\beta$. It is worth noting that in order to
estimate $\beta$ one can choose to measure a {\em finite} number
of $p_k$, i.e., $k=0,\ldots, N-1$. As a matter of fact, this
choice also select a subspace of the Hilbert space where the
reconstructed state is defined.  In turn, (\ref{coh:disp})
corresponds to $N^2$ determinations of the same parameter
$\beta$. Notice that without using the \MinKEnt principle the
only way to exploit the information at disposal, i.e., the
elements of the probability distribution $p_n =
e^{-(\alpha+\beta)^2} (\alpha+\beta)^{2n}/n!$, $\quad
n=0,\ldots,N-1$, is to invert those relations. In order 
to estimate $\beta$ one should solve the set of equations
\begin{equation}\label{coh:disp:no}
-(\alpha + \beta)^2 + 2n\, \ln (\alpha + \beta) =
\ln (\,n!\,p_n),
\end{equation}
which provide only $N$ determinations of $\beta$. 
\section{Conclusions}
\label{s:conc}
In this paper we have considered quantum estimation of states and 
weak Hamiltonian operations in situations where one has at disposal data from the 
measurement of an incomplete set of observables and, at the same time, 
some {\em a priori} information on the state itself. 
By expressing the {\em a priori} information in terms of a bias toward 
a given state the best estimate is obtained using the principle of 
minimum Kullback entropy, i.e., by taking the state that reproduces 
the data while minimizing relative entropy with respect to the bias.
The \MinKEnt principle has been used to estimate the quantum state 
from the measurement of a single observable, either from the sole mean 
value or from the complete probability distribution. In particular,
we have analyzed qubit and harmonic systems with some details.
We have also considered the problem of estimating a {\em weak} 
Hamiltonian processes. In this case there is natural bias of the 
evolved state towards the initial state and the \MinKEnt principle 
can be used as a tool to estimate the Hamiltonian from an incomplete
set of measurements.
\par
Overall the minimum Kullback entropy principle appears to be a convenient
approach for quantum estimation in realistic situations and a useful tool
for the estimation of weak Hamiltonian processes.
\section*{Acknowledgments}
This work was supported by MIUR through the project PRIN-2005024254-002. 
The authors thank A. Ferraro for valuable discussions in the earlier stage of 
this work. MGAP thanks M. Genoni and K. Banaszek for useful discussions.

\end{document}